# VISTA Variables in the Vía Láctea (VVV): Halfway Status and Results

*Maren Hempel[1,2], Dante Minniti[1,3], István Dékány[1], Roberto K. Saito[1,4], Philip W. Lucas[5], Jim Emerson[6], Andrea V. Ahumada[7,8,9], Suzanne Aigrain[10], Maria Victoria Alonso[7], Javier Alonso-García[1], Eduardo B. Amôres[11], Rodolfo Angeloni[1], Julia Arias[12], Reba Bandyopadhyay[13], Rodolfo H. Barbá[12], Beatriz Barbuy[14], Gustavo Baume[15], Juan Carlos Beamin[1], Luigi Bedin[16], Eduardo Bica[17], Jordanka Borissova[18], Leonardo Bronfman[19], Giovanni Carraro[8], Márcio Catelan[1], Juan J. Clariá[7], Carlos Contreras[1], Nicholas Cross[20], Christopher Davis[21], Richard de Grijs[22], Janet E. Drew[5,23], Cecilia Fariña[15], Carlos Feinstein[15], Eduardo Fernández Lajús[15], Stuart Folkes[5,18], Roberto C. Gamen[15], Douglas Geisler[24], Wolfgang Gieren[24], Bertrand Goldman[25], Oscar González[26], Andrew Gosling[27], Guillermo Gunthardt[12], Sebastian Gurovich[7], Nigel C. Hambly[20], Margaret Hanson[28], Melvin Hoare[29], Mike J. Irwin[30], Valentin D. Ivanov[8], Andrés Jordán[1], Eamonn Kerins[31], Karen Kinemuchi[32], Radostin Kurtev[18], Andy Longmore[20], Martin López-Corredoira[33], Tom Maccarone[34], Eduardo Martín[33], Nicola Masetti[35], Ronald E. Mennickent[24], David Merlo[7], Maria Messineo[36], I. Félix Mirabel[37,38], Lorenzo Monaco[8], Christian Moni-Bidin[39], Lorenzo Morelli[40], Nelson Padilla[1], Tali Palma[7], Maria Celeste Parisi[7], Quentin Parker[41,42], Daniela Pavani[17], Pawel Pietrukowicz[43], Grzegorz Pietrzynski[24,44], Giuliano Pignata[45], Marina Rejkuba[8], Alejandra Rojas[1], Alexandre Roman-Lopes[12], Maria Teresa Ruiz[19], Stuart E. Sale[1,18,46], Ivo Saviane[8], Matthias R. Schreiber[18], Anja C. Schröder[47,48], Saurabh Sharma[18], Michael Smith[49], Laerte Sodré Jr.[14], Mario Soto[12], Andrew W. Stephens[50], Motohide Tamura[51], Claus Tappert[18], Mark A. Thompson[5], Ignacio Toledo[52], Elena Valenti[8], Leonardo Vanzi[53], Walter Weidmann[7] and Manuela Zoccali[1]*

(Affiliations can be found after the references)

*The VISTA Variables in the Vía Láctea (VVV) survey is one of six public ESO surveys, and is now in its 4th year of observing. Although far from being complete, the VVV survey has already delivered many results, some directly connected to the intended science goals (detection of variables stars, microlensing events, new star clusters), others concerning more exotic objects, e.g. novae. Now, at the end of the fourth observing period, and comprising roughly 50% of the proposed observations, the actual status of the survey, as well some of the results based on the VVV data, are presented.*

**Introduction**

The Visible and Infrared Survey Telescope for Astronomy (VISTA, Emerson & Sutherland, 2010) has been operated by ESO for four years. Observations began with Science Verification in September 2009, and this was followed by the six public surveys, one of which is VISTA Variables in the Via Láctea (see Minniti et al., 2010; Saito et al., 2010; Catelan et al., 2011). The only science instrument currently available at VISTA is the wide-field camera VIRCAM (Dalton et al., 2006; Emerson & Sutherland, 2010), offering a 1.1 by 1.5 degree field of view, ideal for surveys covering many hundreds of square degrees. The large field of view in combination with the spatial resolution of 0.339 arcseconds per pixel make VISTA/VIRCAM an ideal instrument to observe the most crowded and extincted regions of the Milky Way, i.e. the central regions of the Bulge and the mid-plane regions of the Galactic Disc. The feasible observing period for the VVV survey is limited to six months, between February and October, requiring careful scheduling of the individual observing periods.

The first main phase of the survey, consisting of the YZJHKs multi-colour observations, was assigned to the first semester of the survey (Period 85, 2010) to obtain a first overview of the survey area (described in Section 2). The vast majority of the observations however form the variability campaign, which started in parallel with the multi-colour observations in Period 85, but then occupied all the following observing periods. For the variability campaign, VVV was allotted 300 hours in 2010 (including multi-colour observations), 292 hours in 2011, 275 hours in 2012 and 702 hours in 2013 (split between Periods 90 and 91). The remaining 360 hours of the survey will be used not only to gather additional data for the variability survey (see below), but also to aid the ongoing proper motion study on the Solar Neighbourhood and in searching for microlensing events in a selected Bulge area. The long-term status of the VVV survey, especially in combination with other data (e.g., 2MASS and WISE) make it very suitable to conduct proper motion studies, which are the subject of a later section.

The following observations were scheduled for each period (Minniti et al., 2010):
– P85 (April–October 2010): YZJHKs and additional Ks-band observations for the whole survey area;
– P87 (April–October 2011): Ks-band observations of the complete survey area;
– P89 (April–October 2012): main variability campaign on the Bulge area;
– P90 (November 2012 – March 2013): variability campaign on the outer Disc area;
– P91: (April–October 2013): variability campaign of the inner Disc region and the Bulge.

Already, after the first two observing periods, despite the modest amount of data (see Figure 1) it has become obvious that the data reduction, analysis and even data storage/transfer are daunting tasks. These considerations affected not only the survey team, conducting a wide range of science projects, final quality control, preparation of observations and the public data release, but also the Cambridge Astronomical Survey Unit (CASU), in charge of the pipeline reduction of all data, as well as ESO, performing the observations, the first level of quality control and the final data release.

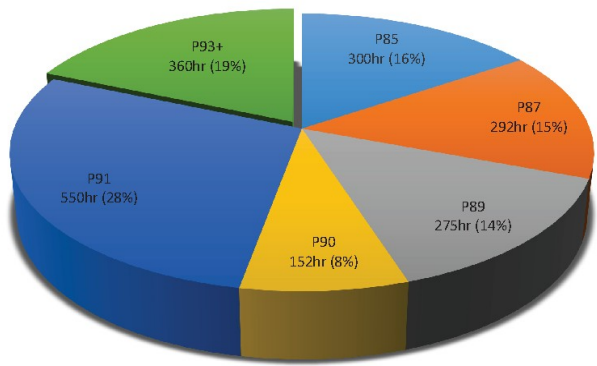

**Figure 1.** Observing schedule for the VVV survey in comparison with the total allotted observing time of 1929 hours. Note, that ca. 50% of the P91 observations are delayed due to weather. The remaining 360 hours will be distributed over Period 93 (2014) and the following years.

## Multi-color photometry

The multi-colour observations in the five broadband filters Y, Z, J, H and Ks commenced in March 2010 and were concluded by September 2011. Starting with the J-, H- and Ks-band data only, an unprecedented first multi-colour view on the inner region of the Milky Way Bulge composed of 84 million stellar sources was presented by Saito et al. (2012a; see also the ESO Release 12421), followed by the study of Soto et al. (2013) of 88 million stellar sources in the southern Galactic Disc.

Multi-colour observations with high spatial resolution and photometric depth, like the VVV datasets, are not only valuable probes of the Galactic structure, but also extremely important to tackle one of the large problems in Galactic studies, the effect of reddening. Line of sight reddening hampers all photometric methods, such as for age and metallicity determination. Detailed reddening maps, such as the ones provided by Schlegel et al. (1998) are widely used to correct for Galactic reddening, but were known to be less reliable for the most reddened regions near the Galactic Plane, where an extinction in the V magnitude (AV) of 30 mag can be reached. Including the single epoch observations in the Y- and Z-bands reveals the distribution of the dust obscuring the inner regions of the Bulge, and enables the high stellar density to become clearly visible (see Figure 2).

Red Clump (RC) stars are an ideal tracer of the reddening effect, since their colour and luminosity are well defined and the dependence on age and metallicity well understood. Gonzales et al. (2013) used the distribution of the RC stars within the VVV Bulge area to derive a reddening map covering the Galactic Bulge (see Figure 3 of Gonzalez et al. [2013]) within 4 degrees of the Galactic Plane with a spatial resolution of 2 arcminutes. At larger Galactic latitudes, where the reddening varies on a larger scale, the resolution is slightly lower, but still ≤ 6 arcminutes. Such a high spatial resolution is, for instance, essential when studying stellar populations in Milky Way star clusters, which, depending on their position, are affected by differential reddening.

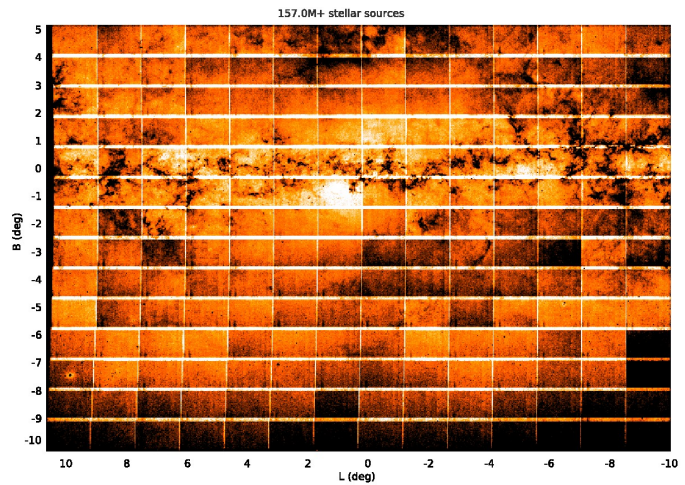

**Figure 2.** The source density in the Bulge region of the VVV survey is shown. Only point sources which were detected in all five bands are included in this plot. Empty fields represent data that did not fulfill the stringent quality criteria and are not part of the public data release. The latter observations will be repeated in due course.

Using the RC stars as bona fide distance indicators of the innermost region of the Bulge along various lines of sight, Gonzalez et al. (2011) could also study the inner Galactic Bar, and provide additional evidence for the existence of a secondary bar structure, as suggested by Nishiyama et al. (2005). In a similar way, Saito et al. (2011), based on 2MASS data, mapped the X-shaped structure of the Bulge, showing extension far beyond the spatial extent of the inner Bar structure.

## Variability in the Near-Infrared

In addition to the multi-colour observations (see above) the first observing period of the survey also included five Ks-band observations for each of the 348 tiles. Although the main variability campaign for the two independent survey areas, i.e. the Bulge and Disc sections, were scheduled for the third and fourth year of VVV, the search for long-term variables required that at least a few epochs for each tile were obtained in each year/ semester. At the time of this article, the Disc area had been observed in the Ks-band 36 times, whereas the Bulge tiles, subject of the observing campaign in P91, had up to 75 epochs available.

In addition, the observations during Science Verification (e.g., Saito et al., 2010) included additional data for tiles b293 and b294 (a total of 71 epochs), coinciding with Baade's Window, which due to its low Galactic reddening had been studied extensively by the Optical Gravitational Lensing Experiment (OGLE) project in the optical, and for which a large number of RR Lyrae stars, the survey's primary targets, are known (see Figure 3).

The backbone of the light curve analysis is the frequency with which the observations were executed. This becomes even more important, when we have to combine data which were not simultaneously observed, as is the case for the individual VVV tiles. In order to be able to compare the derived stellar distances used to create the final 3D model of the inner Milky Way, we have to ensure that the individual light curves are indeed of comparable quality, i.e., that the light curves for the variable stars are populated in a similar fashion. As for any other observing campaign, this not only depends on the original schedule (e.g., Minniti et al., 2010; Saito et al., 2012b), but is also affected by the observing conditions, e.g., presence of clouds, high humidity, etc. As a result, the VVV observations vary significantly in their cadence, as shown in Figure 4.

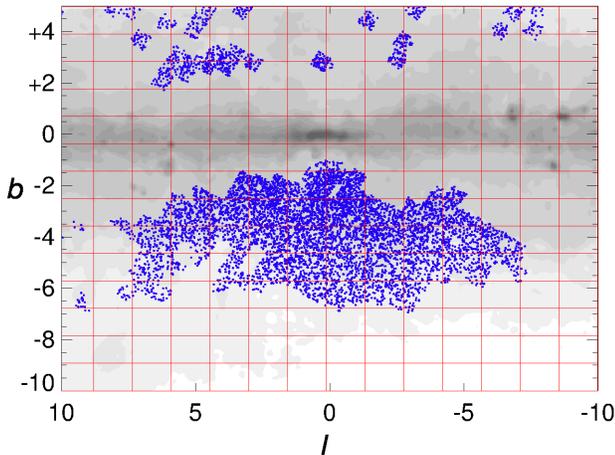

**Figure 3.** The bulge region of the VVV survey (196 tiles, red boxes). The blue dots mark the position of known RR Lyrae (> 13000), based on the OGLE data base.

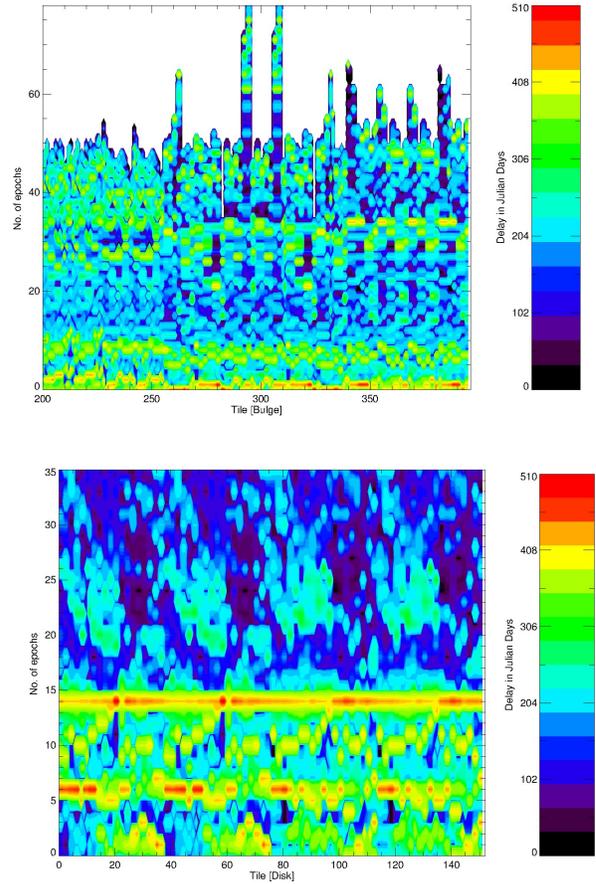

**Figure 4.** Distribution of the delays between two consecutive epochs of individual $K_s$-band observations for the Bulge (top panel) and the Disk region (bottom panel) of the VVV survey. The timeline covered by the observations is February 2010 to October 2013.

Although the observations are still ongoing and will require an additional 2–3 years, the analysis of the light curves has already begun; they show the excellent quality of the data. As shown in Figures 5 and 6, even at this early stage of the survey, and based on a limited number of observing epochs, long-term variables with periods of several hundreds of days (Figure 5) can be detected as unambiguously as well as the objects with a period of only a few days or even hours (Figure 6). Only surveys with an anticipated duration period of many years are suitable to study long-term variables such as the ones shown in Figure 5.

Although the VVV survey is targeted at various types of variable stars, of special importance are those classes of variable stars that show a well-defined luminosity–period correlation, because they will allow us to derive distances and eventually build a three-dimensional model of the observed Milky Way region. In addition the frequent observations, in combination with a limiting Ks magnitude of ~18.0 for a single epoch, allows us to search for and follow up more exotic objects, like novae (Beamin et al., 2013; Saito et al. 2013.

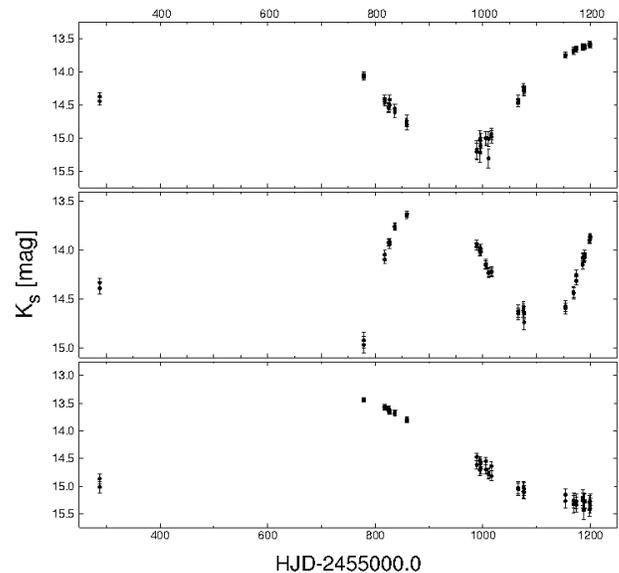

**Figure 5.** Examples of Ks-band light curves of long term variables. The light curves are based on the aperture photometry provided by the CASU VIRCAM pipeline, and use the individual pawprints (Emerson et al., 2004).

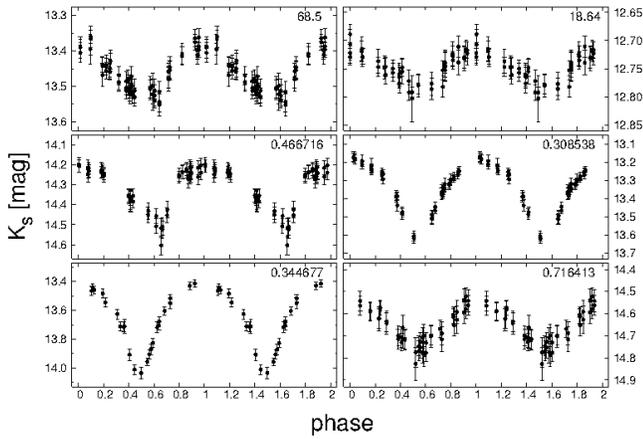

**Figure 6.** Examples of six Ks-band light curves of variable stars with periods ranging from a few hours up to several weeks are displayed. As in Figure 5 the aperture photometry of the CASU source catalogues was used.

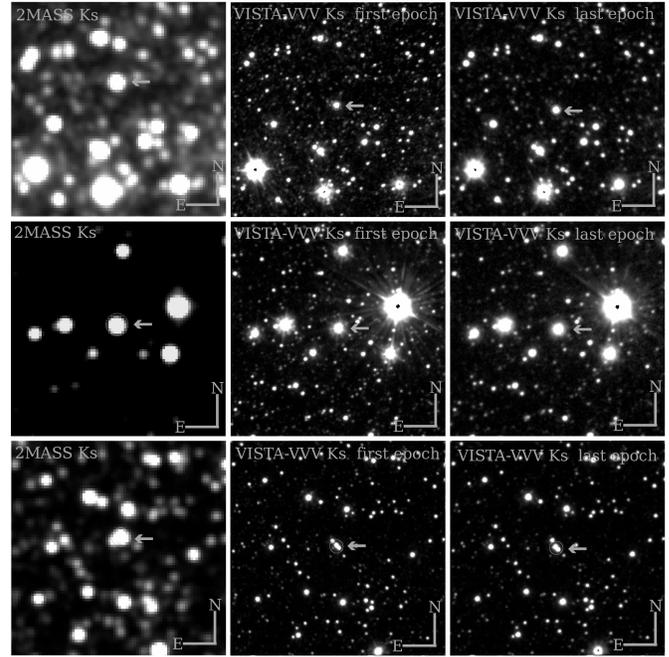

**Figure 7.** Selected sample of HPM objects, detected by Gromadzki et al. (2013, see also their Figure 1). The left column shows the 2MASS images, whereas the middle and right columns show the first and last VVV image taken of the object, used to derive the proper motions. Shown from top to bottom are: an HPM M dwarf; an M dwarf + white dwarf common proper motion binary; and a close common proper motion binary.

### Proper motions

The above-mentioned delays in the observing schedule, which are anything but helpful for the general progress of the survey, can be considered as a blessing in disguise for the proper motion studies, another essential part of the survey (Minniti et al., 2010). Proper motions are the most widely used method to search for close Solar neighbours and thus to complete the census of stars within a few tens of parsecs from the Sun. In particular, in the direction towards the Milky Way ulge, severe crowding and extinction hamper the detection of faint, albeit nearby stars. (e.g., Lépine et al., 2008); hese drawbacks can however be vercome y the VVV observations. The 0.339-arcsecond pixels, the much reduced dust extinction (in comparison with optical observations) and the long time-line of (eventually) up to seven years (i.e., the full duration of the VVV survey) make this survey an ideal tool with which to find those missing nearby stars. Already at this early stage of the survey, Gromadzki et al. (2013) have found several hundreds of those elusive objects (Figure 7) and, even more importantly, while searching only about 31% of the survey area, which corresponds to ~1% of the sky.

The VVV catalogue of high proper motion (HPM) objects includes M dwarfs towards the Galactic Bulge, common proper motion binaries (see also Ivanov et al., 2013), close common proper motion M dwarfs + white dwarf binaries and brown dwarfs towards the Galactic Bulge. At this point of the survey proper motions of ~1 arcseconds/yr are studied. Once the late phase of the survey is added, and also the earlier observations with 2MASS (Skrutskie et al., 2006) are included for the brighter ($K_s < 14$ mag) objects, an accuracy in the proper motion measurement that almost rivals the one of the Gaia mission (i.e. proper motion error: ~ 10 micro-arcseconds/yr) will be achievable. Further, near-infrared observations as provided by VVV, 2MASS and DENIS (e.g., Epchtein et al., 1994) allow us to carry out these studies in an area inaccessible for optical missions, such as Gaia.

### The Future

In recent years, 2MASS has become a veritable treasure trove for Milky Way studies, and the VVV survey will undoubtedly be its worthy successor for the innermost regions of the Milky Way. Extending the source lists by applying more sophisticated detection algorithms, such as point spread function photometry, will allow fainter sources to be detected and hence allow distant objects at the far side of the Galaxy to be probed. The distances derived for the variable stars within the VVV survey area, together with the proper motions, will allow us to build a model of the mostly unexplored inner regions of the Galaxy. By approximately 2016, the observations for the VVV survey will be complete and provide the data required to build a detailed model of the Milky Way Bulge and the southern Galactic Disc. From there we can take a large step closer to understanding the structure and dynamics of the Galaxy.

### Acknowledgements

The VVV Survey is supported by ESO, the BASAL Center for Astrophysics and Associated Technologies PFB-06, the FONDAP Center for Astrophysics 15010003 and the MIDEPLAN Milky Way Millennium Nucleus P07-021-F. We would like to thank the staff of CASU and WFAU, who provide us with the pipeline processing, data calibration and archive. Some VVV tiles were made using the Aladin sky atlas, SExtractor software and products from TERAPIX pipeline. This publication makes use of data products from the Two Micron All Sky Survey, a joint project of the University of Massachusetts and IPAC/CALTECH, funded by NASA and the NSF.

[1] Instituto de Astrofísica, Pontificia Universidad Católica de Chile, Av. Vicuña Mackenna 4860, Casilla 360, Santiago 22, Chile

[2] The Millennium Institute of Astrophysics, Av. Vicuña Mackenna 4860, Casilla 360, Santiago 22, Chile 782-0436 Macul, Santiago, Chile

[3] Vatican Observatory, Vatican City State V-00120, Italy

[4] Universidade Federal de Sergipe, Departamento de Física, Av. Marechal Rondon s/n, 49100-000, São Cristóvão, SE, Brazil

[5] Centre for Astrophysics Research, Science and Technology Research Institute, University of Hertfordshire, Hatfield AL10 9AB, UK

[6] Astronomy Unit, School of Mathematical Sciences, Queen Mary, University of London, Mile End Road, London, E1 4NS, UK

[7] Observatorio Astronómico de Córdoba, Universidad Nacional de Córdoba, Laprida 854, 5000 Córdoba, Argentina

[8] European Southern Observatory, Av. Alonso de Córdova 3107, Casilla 19, Santiago 19001, Chile

[9] Consejo Nacional de Investigaciones Científicas y Técnicas, Av. Rivadavia 1917 - CPC1033AAJ - Buenos Aires, Argentina

[10] School of Physics, University of Exeter, Exeter EX4 4QL; Oxford Astrophysics, University of Oxford, Keble Road, Oxford OX1 3RH, UK

[11] SIM, Faculdade de Ciências da Universidade de Lisboa, Ed. C8, Campo Grande, 1749-016, Lisboa, Portugal

[12] Departamento de Física, Universidad de La Serena, Benavente 980, La Serena, Chile

[13] Department of Astronomy, University of Florida, 211 Bryant Space Science Center P.O. Box 112055, Gainesville, FL, 32611-2055, USA

[14] Universidade de São Paulo, IAG, Rua do Matão 1226, Cidade Universitária, São Paulo 05508-900, Brazil

[15] Facultad de Ciencias Astronómicas y Geofísicas, Universidad Nacional de La Plata, and Instituto de Astrofísica La Plata, Paseo del Bosque S/N, B1900FWA, La Plata, Argentina

[16] Space Telescope Science Institute, 3700 San Martin Drive, Baltimore, MD 21218, USA

[17] Universidade Federal do Rio Grande do Sul, IF, CP 15051, Porto Alegre 91501-970, RS, Brazil

[18] Departamento de Física y Astronomía, Facultad de Ciencias, Universidad de Valparaíso, Ave. Gran Bretaña 1111, Playa Ancha, Casilla 5030, Valparaíso, Chile

[19] Departamento de Astronomía, Universidad de Chile, Casilla 36-D, Santiago, Chile

[20] Institute for Astronomy, The University of Edinburgh, Royal Observatory, Blackford Hill, Edinburgh EH9 3HJ, UK

[21] Joint Astronomy Centre, 660 North A'ohōkū Place, University Park, Hilo, HI 96720, USA

[22] Kavli Institute for Astronomy and Astrophysics, Peking University, Yi He Yuan Lu 5, Hai Dian District, Beijing 100871, China

[23] Astrophysics Group, Imperial College London, Blackett Laboratory, Prince Consort Road, London SW7 2AZ, UK

[24] Departmento de Astronomía, Universidad de Concepción, Casilla 160-C, Concepción, Chile

[25] Max Planck Institute for Astronomy, Königstuhl 17, 69117 Heidelberg, Germany

[26] European Southern Observatory, Kar-Schwarzschild Str. 2, 85748 Garching, Germany

[27] Department of Astrophysics, University of Oxford, Keble Road, Oxford OX1 3RH

[28] Department of Physics, University of Cincinnati, Cincinnati, OH 45221-0011, USA

[29] School of Physics & Astronomy, University of Leeds, Woodhouse Lane, Leeds LS2 9JT, UK

[30] Institute of Astronomy, University of Cambridge, Madingley Road, Cambridge CB3 0HA, UK

[31] Jodrell Bank Centre for Astrophysics, The University of Manchester, Oxford Road, Manchester M13 9PL, UK

[32] NASA-Ames Research Center, Mail Stop 244-30 Moffett Field, CA 94035, USA

[33] Instituto de Astrofísica de Canarias, Vía Láctea s/n, E38205 - La Laguna (Tenerife), Spain

[34] Department of Physics, Texas Tech University, Box 41051, Lubbock, TX 79409-1051, USA

[35] Istituto di Astrofisica Spaziale e Fisica Cosmica di Bologna, via Gobetti 101, 40129 Bologna, Italy

[36] Rochester Institute of Technology, One Lomb Memorial Drive, Rochester, NY 14623, USA

[37] Service d'Astrophysique - IRFU, CEA-Saclay, 91191 Gif sur Yvette, France

[38] Instituto de Astronomía y Física del Espacio, Casilla de Correo 67, Sucursal 28, Buenos Aires, Argentina

[39] Instituto de Astronomía, Universidad Católica del Norte, Avenida Angamos 0610, Antofagasta, Chile

[40] Dipartimento di Astronomia, Universitá di Padova, vicolo dell'Osservatorio 3, 35122 Padova, Italy

[41] Department of Physics, Macquarie University, Sydney, NSW 2109, Australia

[42] Anglo-Australian Observatory, PO Box 296, Epping, NSW 1710, Australia

[43] Nicolaus Copernicus Astronomical Center, ul. Bartycka 18, 00-716 Warsaw, Poland

[44] Warsaw University Observatory, Al. Ujazdowskie 4,00-478, Warsaw, Poland

[45] Departamento de Ciencias Físicas, Universidad Andres Bello, Av. República 252, Santiago, Chile

[46] Rudolf Peierls Centre for Theoretical Physics, Keble Road, Oxford, OX1 3NP, UK

[47] SKA/KAT, Lonsdale Building, Lonsdale Road, Pinelands 7405, Cape Town, South Africa

[48] Hartebeesthoek Radio Astronomy Observatory, PO Box 443, Krugersdorp 1740, South Africa

[49] The University of Kent, Canterbury, Kent, CT2 7NZ, UK

[50] Division of Optical and Infrared Astronomy, National Astronomical Observatory of Japan 2-21-1 Osawa, Mitaka, Tokyo, 181-8588, Japan

[51] Gemini Observatory, Northern Operations Center, 670 N. A'ohoku Place, Hilo, Hawaii, 96720, USA

[52] ALMA Observatory, Alonso de Córdova 3107, Vitacura, Santiago,Chile

[53] Departamento de Ingeniería Eléctrica, Pontificia Universidad Católica de Chile, Av. Vicuña Mackenna 4860, Santiago, Chile